\RequirePackage{fix-cm}
\documentclass[twocolumn]{svjour3}          
\smartqed  
\usepackage{graphicx,amsmath,amssymb,color}
 \usepackage{mathptmx}   
\usepackage[T1]{fontenc}
\usepackage[utf8]{inputenc}
\newcommand{\cn}{\,{\sf cn}}
\newcommand{\sn}{\,{\sf sn}}
\newcommand{\dn}{\,{\sf dn}}

 \journalname{Nonlinear Dynamics}
\begin{document}

\title{Superposition solutions to the extended KdV equation for water surface waves 
}

\titlerunning{Superposition solutions }    

\author{ Piotr Rozmej  \and  Anna Karczewska \and  Eryk Infeld}
 

\institute{P. Rozmej \at Faculty of Physics and Astronomy,
University of Zielona G\'ora, Szafrana 4a, 65-246 Zielona G\'ora, Poland \\
              Tel.: +48-68-3282909\\
              Fax: +48-68-3282920\\
              \email{P.Rozmej@if.uz.zgora.pl}           
           \and
           A. Karczewska \at Faculty of Mathematics, Computer Science and Econometrics, University of Zielona G\'ora, Szafrana 4a, 65-246 Zielona G\'ora, Poland\\
              \email{A.Karczewska@wmie.uz.zgora.pl}           
            \and 
            E. Infeld \at  National Centre for Nuclear Research, Hoża 69, 00-681 Warszawa, Poland\\ 
      \email{Eryk.Infeld@ncbj.gov.pl}
}

\date{Received: \today / Accepted: }

\maketitle

\begin{abstract}
The KdV equation can be derived in the shallow water limit of the Euler equations. Over the last few decades, this equation has been extended to include higher order effects. Although this equation has only one conservation law, exact periodic  and solitonic solutions exist. 
Khare and Saxena  \cite{KhSa,KhSa14,KhSa15} demonstrated the possibility of generating new
exact solutions by combining known ones for several fundamental
equations (e.g., Korteweg - de Vries, Nonlinear Schr\"{o}dinger). Here
we find that this construction can be repeated for higher order, non-integrable extensions of these equations. Contrary to many statements
in the literature, there seems to be no correlation between integrability and the number of nonlinear one variable wave solutions.

\end{abstract}

\section{Introduction} \label{intro}

A long time ago, Stokes opened the
field of nonlinear hydrodynamics by showing that waves
described by nonlinear models can be periodic \cite{Stokes}. Although several related results followed, it took half a century before the Korteweg - de Vries equation became widely known \cite{KdV}.
A more accurate equation system, Boussinesq, was formulated
in 1871. 
It is also the theme of several recent papers \cite{BBM,Bona81}.
Another direction research
has gone in is including perpendicular dynamics in KdV, e.g., \cite{IRS99}.

The KdV equation is one of the most succesful physical equations. It consists of the mathematically simplest possible terms representing the interplay of nonlinearity and dispersion. This simplicity may be one of the reasons for success. Here we investigate this equation, improved as derived  from the Euler inviscid and irrotational water equations. 

Just as for conventional KdV, two small parameters are assumed: wave amplitude/depth $(a/H)$ and depth/wave\-length squared $(H/l)^2$. These dimensionless expansion constants are called $\alpha$ and $\beta$. We take the expansion one order higher. The new terms will then be of second order. This procedure limits considerations to waves for which the two parameters are comparable. Unfortunately some authors tend to be careless about this limitation.

The next approximation to Euler's equations for long waves over a shallow riverbed is ($\eta$ is the elevation above a flat surface divided by $H$)
\begin{align} \label{kdv2}
\eta_t & +  \eta_x + \frac{3}{2} \alpha\,\eta\eta_x+ \frac{1}{6}\beta\, \eta_{3x} -\frac{3}{8}\alpha^2\eta^2\eta_x \\ & +
  \alpha\beta\,\left(\frac{23}{24}\eta_x\eta_{2x}+\frac{5}{12}\eta\eta_{3x} \right)+\frac{19}{360}\beta^2\eta_{5x} =0 .  \nonumber
\end{align}
In (\ref{kdv2}) and subsequently we use low indexes for derivatives $\left(\eta_{nx}\equiv\frac{\partial^n \eta}{\partial x^n} \right)$. 
This second order equation was called by \linebreak Marchant and Smyth \cite{MS90,MS96} the \textit{extended KdV}. It was also derived in a different way in \cite{BS13} and \cite{KRR14,KRI14}. We call it {\bf KdV2}. It is not integrable. However, by keeping the same terms but changing one numerical coefficient (specifically, replacing $\frac{23}{24}$ by $\frac{5}{6}$) we can obtain an integrable equation \cite{KRI15,Kalisch}.

Not only  is KdV2 non-integrable, it only seems to have one conservation law (volume or mass) \cite{SV16}. However a simple derivation of adiabatically conserved quantities can be found in \cite{KRIR17}.

Recently, Khare and Saxena \cite{KhSa,KhSa14,KhSa15} demonstrated that for several nonlinear equations which admit solutions in terms of elliptic functions $\cn(x,m),\dn(x,m)$ there exist solutions in terms of superpositions $\cn(x,m)\pm \sqrt{m}\dn(x,m)$. They also showed that KdV which admits solutions in terms of $\dn^2(x,m)$ also admits solutions in terms of superpositions $\dn^2(x,m)\pm \sqrt{m}\cn(x,m)\dn(x,m)$. Since then we  found analytic solutions to KdV2 in terms of $\cn^2(x,m)$ \cite{IKRR17a,IKRR17b} the results of Khare and Saxena \cite{KhSa,KhSa14,KhSa15} inspired us to look for solutions to KdV2 in similar form.

\section{Exact periodic solutions for KdV2}\label{sec1}

First, we repeat shortly the results obtained by Khare and Saxena \cite{KhSa}, but formulating them for KdV in a fixed frame, that is, for the equation
\begin{equation} \label{kdv}
\eta_t  +  \eta_x + \frac{3}{2} \alpha\,\eta\eta_x+ \frac{1}{6}\beta\, \eta_{3x}  =0 .  
\end{equation}
Assuming solution in the form 
\begin{equation} \label{adn2}
\eta(x,t) = A\dn^2 [B(x-vt),m]
\end{equation}
one finds 
$$A= \frac{4}{3}\,\frac{ B^2 \beta}{\alpha}\quad \mbox{and} \quad v=1+\frac{2}{3}\beta B^2(2-m)= 1+\frac{\alpha}{2}A(2-m).$$
Next, the authors \cite{KhSa} showed that superpositions 
\begin{align} \label{adncn}
\eta_{\pm}(x,t) & = \frac{1}{2} A \left(\dn^2 [B(x-vt),m] \right. \\
& \hspace{6ex} \left. \pm \sqrt{m} \cn [B(x-vt),m]\dn [B(x-vt),m] \right) \nonumber 
\end{align}
are  solutions to (\ref{kdv}) with the same relation between $A$ and $B$, but for a different velocity, ~$v_{\pm} =1+\frac{1}{6}\beta B^2(5-m)$.

Now, we look for periodic nonlinear wave solutions of KdV2 (\ref{kdv2}). 
Introduce $y:=x-v t$. Then 
 $\eta(x,t) =\eta(y)$, \linebreak $\eta_t=-v\eta_y$~ and equation (\ref{kdv2}) takes the form of an ODE
\begin{align} \label{kdv2y}
(1-v) \eta_y & + \frac{3}{2} \alpha\,\eta\eta_y+ \frac{1}{6}\beta\, \eta_{3y} -\frac{3}{8}\alpha^2\eta^2\eta_y \\ & +
  \alpha\beta\,\left(\frac{23}{24}\eta_y\eta_{2y}+\frac{5}{12}\eta\eta_{3y} \right)+\frac{19}{360}\beta^2\eta_{5y} =0 .  \nonumber
\end{align}

\subsection{Single periodic function $\dn^2$} \label{ss21}

First, we recall some properties of the Jacobi elliptic functions (arguments are omitted)
\begin{equation} \label{jac}
\sn^2+\cn^2 =1, \quad \dn^2+m\sn^2 = 1.
\end{equation}
Their derivatives are
\begin{equation} \label{jder}
\frac{d\,\sn}{dy} =  \cn\,\dn,\quad \frac{d\,\cn}{dy} = - \sn\,\dn, \quad  \frac{d\,\dn}{dy} = -m \,\sn\,\cn .
\end{equation}

Assume a solution of (\ref{kdv2}) in the same form as KdV solution (\ref{adn2}). Insertion of (\ref{adn2}) into (\ref{kdv2y}) yields 
\begin{equation} \label{rd1}
\frac{ABm}{180} \cn\dn\sn \left(F_0+F_2 \cn^2 + F_4 \cn^4  \right)=0.
\end{equation}
Equation (\ref{rd1}) holds for arbitrary arguments when $F_0,F_2,F_4$ vanish simultaneously. The explicit form of this set of equations is following
\begin{align} \label{f0}
F_0 & = 135 \alpha ^2 A^2 (m\!-\!1)^2\!+\!30 \alpha  A
   (m\!-\!1) \left(\beta  B^2 (63
   m\!-\!20)\!+\!18\right)  \nonumber  \\ & \hspace{2ex}
 -8 \left(19 \beta ^2 B^4
   \left(17 m^2\!-\!17 m+2\right)+30 \beta  B^2
   (2 m\!-\!1)+45\right) \nonumber \\ & \hspace{2ex}
+360 v = 0,  \\ \label{f2}
F_2 & =  -30 m \left[ 9 \alpha ^2 A^2 (m\!-\!1)+6 \alpha 
   A \left(\beta  B^2 (32 m\!-\!21)+3\right) \right. \nonumber \\ 
   & \hspace{2ex} \left. -8
   \beta  B^2 \left(19  \beta  B^2 (2
   m-1)+3\right)\right]  = 0, \\ \label{f4}
F_4 & = 45 m^2 \left(3 \alpha ^2 A^2+86 \alpha  A
   \beta  B^2-152 \beta ^2 B^4\right) = 0.
\end{align}
Equation (\ref{f4}) is equivalent to the  \cite[Eq.~(26)]{KRI14} obtained for solitonic solutions to KdV2. 
Denoting ~$z:=\frac{B^2\beta}{A\,\alpha}$ one obtains from (\ref{f4}) two possible solutions 
\begin{equation} \label{z12}
 z_1=\frac{43-\sqrt{2305}}{152} <0 \quad \mbox{and} \quad z_2=\frac{43+\sqrt{2305}}{152}>0 \,.
\end{equation}
\textcolor{blue}{ 
The case $z=z_1$ leads to $B^2<0$ and has to be rejected as in previous papers \cite{KRI14,IKRR17a}.
Then for $z=z_2 $ the amplitude $A$ is
\begin{align}  \label{A12}
  A &= \frac{43+\sqrt{2305}}{3}\,\frac{B^2\beta}{\alpha}  >0 \,.
\end{align} 
Inserting this into (\ref{f2}) yields
\begin{align} \label{B12}
 B^{2} & = \frac{3 \left(\sqrt{2305}-14\right)}{703 \beta  (2-m)}
\end{align}
and then using (\ref{A12}) one has
\begin{align}  \label{A2}
  A &= \frac{3 \left(\sqrt{2305}-51\right)}{37 \alpha  (m-2)} .
\end{align} 
}

\textcolor{blue}{
Finally from (\ref{f0}) one obtains 
\begin{align} \label{vv1}
 v & =\left\{  4 \left(129877+314 \sqrt{2305}\right) m^2 
\right.\nonumber. \\ & \left. \hspace{3ex} 
+\left(18409 \sqrt{2305}-3209623\right)(m-1)\right\} \nonumber.\\ & \hspace{3ex}\left/ \hspace{1ex} \left\{ 520220 (m-2)^2\right\}\right. .
\end{align}
}

Despite the same form of solutions to KdV and KdV2  
there is a fundamental difference.  KdV only imposes two conditions on coefficients $A,B,v$ for given $m$, so there is one parameter freedom.
This is no longer the case for KdV2.\\

\begin{figure}[tbh]
\begin{center}
\resizebox{0.99\columnwidth}{!}{\includegraphics{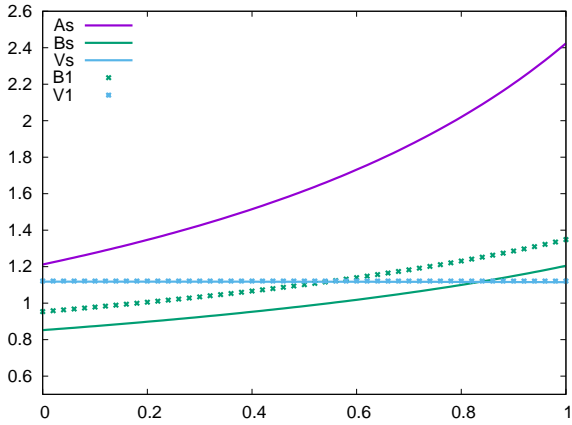}}
\end{center}
\caption{Coefficients $A, B, v$, $(z=z_2)$ as functions of $m$ for the periodic solution of KdV2 in the form of a single function (\ref{adn2}). Lines represent KdV2 coefficients, points  KdV coefficients.  ($A_{\text{KdV}}=A_{\text{KdV2}}$).} 
\label{ABVsingle}\end{figure}

\noindent {\bf Comparison to KdV solutions}\\
Is a solution of KdV2  much different from the KdV solution for the same $m$? In order to compare solutions of both equations, remember that the set of three equations (\ref{f0})-(\ref{f4}) fixes all $A,B,v$ coefficients for KdV2 for given $m$. In the case of KdV the equation  analogous to (\ref{rd1}) only imposes  two conditions on three parameters.
Therefore one parameter, say amplitude $A$, can be chosen arbitrary.
Then we compare coefficients of solutions to KdV2 and KdV chosing the same 
value of $A$, that is, $A_{\text{KdV2}}$. Such comparison is displayed in Fig.~\ref{ABVsingle} for $\alpha=\beta=\frac{1}{10}$.

It is clear that $v_{\text{KdV2}}$ and $v_{\text{KdV}}$ are very similar.  We have the following relations: for KdV $\frac{B^2}{A}=\frac{3\,\alpha}{4\,\beta}$, whereas for KdV2 $\frac{B^2}{A}=\frac{\alpha}{\beta}z_2$. Since $z_2\approx 0.6$, $B_{\text{KdV}}/B_{\text{KdV}}=\sqrt{\frac{3~}{4 z_2}}\approx 1.12$. The same relations hold between KdV2 and KdV coefficients for superposition solutions shown in Fig.~\ref{ABVcndn}.

\textcolor{blue}{
The above examples for the case $\alpha=\beta=\frac{1}{10}$ show that for somewhat small values of $\alpha$ the coefficients of KdV2 $\dn^2$ solutions are not much different from those of KdV.  }

\textcolor{blue}{
However, physically relevant exact solutions of  $\dn^2$ to KdV2 can be found for much larger values of the parameter $\alpha$. In figure 2 
the amplitude $A(\alpha,m)$ given by (\ref{A2}) is shown as a contour plot for the region $\alpha\in [0.01,0.5], \hspace{1ex} m\in[0,1]$. It is clear that reasonable amplitudes occur in  wide regions of $\alpha$ and $m$ values. The amplitudes become too big only for  $\alpha\to 0$, but in such cases KdV works very well.
\begin{figure}[tbh]
\begin{center} 
\resizebox{0.99\columnwidth}{!}{\includegraphics{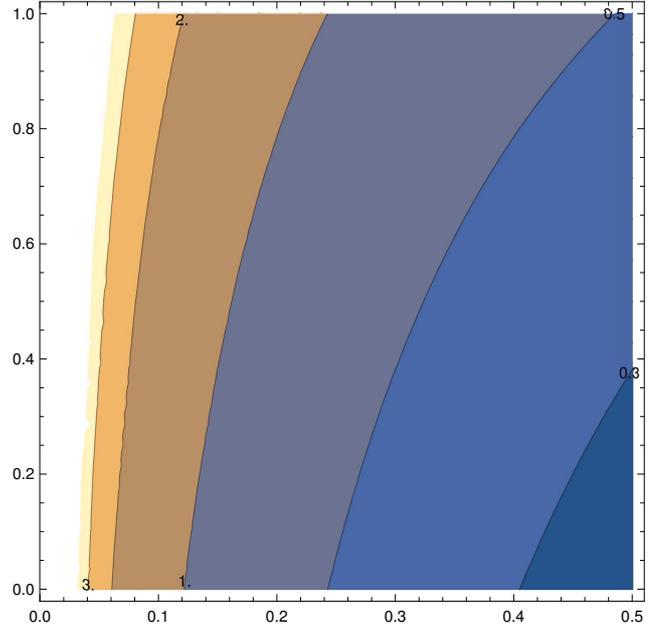}}
\end{center}\textcolor{blue}{
\caption{The amplitude $A(\alpha,m)$ given by (\ref{A2}) as function of $(\alpha,m)$.} }
\label{Kontur1}\end{figure}
}

\subsection{Superposition "$\dn^2+\sqrt{m}\cn\dn$"} \label{ss22}

Now assume the periodic solution to be in the same form as the corresponding  superposition solution of KdV  \cite{KhSa} functio
\begin{equation} \label{ey}
 \eta_{+}(y) = \frac{1}{2} A \left[ \dn^2(By,m) +\sqrt{m}\,\cn(By,m)\dn(By,m) \right] ,
\end{equation}
where $A,B,v$ are yet unknown constants ($m$ is the elliptic parameter).  
We will need 
\begin{align} \label{y1}
\eta_y~ &= -\frac{1}{2}AB\,\sqrt{m} \left(\sqrt{m}\cn +\dn\right)^2\sn, \\ \label{y2a}
\eta_{2y} &= \frac{1}{2}AB^2\,\sqrt{m}\left(\sqrt{m}\cn +\dn\right)^2\!\left(\!-\!\cn\dn\!+\!2\sqrt{m}\sn^2\right),\\ \label{y3a}
\eta_{3y} &=  \frac{1}{2}AB^3\,\sqrt{m}\left(\sqrt{m}\cn +\dn\right)^2  \\ & \hspace{2ex} \times 
\sn\, \left(m\cn^2+6\sqrt{m}\cn\dn+\dn^2-4m\sn^2\right)
,\nonumber\\ \label{y5a}
\eta_{5y} &= -\frac{1}{2} AB^5 \,\sqrt{m}\left(\sqrt{m}\cn +\dn\right)^2 \\ &  \hspace{2ex} \times 
\sn \left[m^2\cn^4+30 m^{3/2}\cn^3\dn +\dn^4 -44\dn^2\sn^2\right. \nonumber \\ &  \hspace{7ex} 
+16m^2\sn^4  -30\sqrt{m}\cn\dn\left(-\dn^2+4m\sn^2\right) \nonumber \\&  \hspace{7ex} \left.
+ \cn^2\left(74m\dn^2-44m^2\sn^2\right)\right] .\nonumber
\end{align}

Denote (\ref{kdv2y}) as
\begin{equation} \label{eq2}
E_1+E_2+E_3+E_4+E_5+E_6+E_7=0,
\end{equation}
where
\begin{align} \label{E1}
E_1 &= (1\!-\!v) \eta_y =-\frac{1}{2}AB(1\!-\!v)\sqrt{m} \left(\sqrt{m}\cn \!+\!\dn\right)^2\!\sn, \\ \label{E2}
E_2 &= \frac{3}{2} \alpha\,\eta\eta_y =-\frac{3}{8} \alpha\, A^2B\,\sqrt{m} \left(\sqrt{m}\cn +\dn\right)^3\sn\dn, 
\\ \label{E3}
E_3 &= \frac{1}{6}\beta\eta{3y}= \frac{1}{12}\beta\, A B^3 \,\sqrt{m}\left(\sqrt{m}\cn +\dn\right)^2 \sn \\ & \hspace{10ex} 
\left(m\cn^2+6\sqrt{m}\cn\dn+\dn^2-4m\sn^2  \right), \nonumber
\\ \label{E4}
E_4 &= -\frac{3}{8}\alpha^2\eta^2\eta_y = \frac{3}{64}\alpha^2 A^3B \sqrt{m}\dn^2\!\left(\sqrt{m}\cn\! +\!\dn\right)^4\!\sn , 
\\ \label{E5}
E_5 &= \frac{23}{24}\alpha\beta\,\eta_y\eta_{2y}=-\frac{23}{96}\alpha\beta\,A^2B^3 \,m\,\left(\sqrt{m}\cn +\dn\right)^4 \sn \nonumber  \\
 & \hspace{10ex} 
 \left(-\cn\dn+2 \sqrt{m}\sn^2 \right),  \\ \label{E6}
 E_6 &= \frac{5}{12}\alpha\beta\,\eta\eta_{3y} = \frac{5}{48}\alpha\beta\,A^2B^3\, \sqrt{m}\dn\left(\sqrt{m}\cn +\dn\right)^3 \sn \nonumber \\
 & \hspace{10ex} 
\left(m\cn^2+6 \sqrt{m}\cn\dn+\dn^2-4m\sn^2 \right) , 
 \\ \label{E7}
 E_7&=\frac{19}{360}\beta^2\eta_{5y} = -\frac{19}{720}\beta^2 A B^5  \,\sqrt{m}\left(\sqrt{m}\cn +\dn\right)^2\sn \nonumber \\
 & 
\left[m^2\cn^4+30 m^{3/2}\cn^3\dn +\dn^4-44\dn^2\sn^2 \right.\nonumber \\ 
& \hspace{2ex}
+16m^2\sn^4 -30\sqrt{m}\cn\dn\left(-\dn^2+4m\sn^2\right) \\ & \hspace{2ex}  \left. 
+ \cn^2\left(74m\dn^2-44m^2\sn^2\right)\right] .\nonumber
 \end{align}
Then (\ref{eq2}) becomes 
\begin{align}  \label{eq2a}
  \frac{1}{2} AB\,\sqrt{m} \left(\sqrt{m}\cn +\dn\right)^2\sn \hspace{20ex} & \\ \hspace{1ex} \times
\left( F_0 +F_{cd} \cn\dn + F_{c^2} \cn^2 + F_{c^3d}\cn^3\dn+ F_{c^4} \cn^4\right) & =0. \nonumber
\end{align}
 
Equation (\ref{eq2a}) is valid for arbitrary arguments  when all  coefficients $F_0,F_{cd},F_{c^2},F_{c^3d}, F_{c^4}$ vanish simultaneously. This gives us a set of  equations for the coefficients $A,B,v$
\begin{align} \label{c1}
F_0 & =-1440 v-135 \alpha ^2 A^2 (m-1)^2 \nonumber \\ & \hspace{2ex}
 -60 \alpha  A (m-1) \left[\beta  B^2 (48 m-5) +18\right] \\ & \hspace{2ex}
 +4 \left[19 \beta^2 B^4 \left(61 m^2-46 m+1\right) \right.\nonumber  \\ & \hspace{6ex} \left.
   +60 \beta  B^2 (5 m-1)+360\right] =0, \nonumber \\ \label{c2}
F_{cd} &= 30 \sqrt{m} \left[ 9 \alpha ^2 A^2 (m-1)+3 \alpha  A
   \left(\beta  B^2 (75 m-31)+12\right)\right.\nonumber  \\ & \hspace{8ex} \left.
-4 \beta  B^2
   \left(19 \beta  B^2 (5 m-1)+12\right)\right]   =0,
\\ \label{c3}
F_{c^2} & =  15 m \left(27 \alpha ^2 A^2 (m-1)+12 \alpha  A
   \left(\beta  B^2 (59 m-37)+6\right)\right.\nonumber  \\ & \hspace{8ex} \left.
-32 \beta  B^2
   \left(19 \beta  B^2 (2 m-1)+3\right)\right)
 =0 ,\\ \label{c4}
F_{c^3d} & = -90 m^{3/2} \left(3 \alpha ^2 A^2\!+\!86 \alpha  A \beta 
   B^2\!-\!152 \beta ^2 B^4\right) 
 =0 ,\\ \label{c5}
F_{c^4} & =  -90 m^2 \left(3 \alpha ^2 A^2+86 \alpha  A \beta 
   B^2-152 \beta ^2 B^4\right)
 =0 .
 \end{align}

Equations (\ref{c4}) and (\ref{c5}) are equivalent and give the same condition as (\ref{f4}).
Solving (\ref{c4}) with respect to $B^2$, we obtain the same relations as in \cite[Eq.~(28)]{KRI14}
\begin{equation} \label{bb}
(B_{1/2})^2= \frac{A\alpha}{ \beta}\left(\frac{43\mp \sqrt{2305}}{152}\right)
\end{equation}
Denote 
\begin{equation} \label{zz}
 z_1=\frac{43-\sqrt{2305}}{152} \qquad \mbox{and} \qquad z_2=\frac{43+\sqrt{2305}}{152} \,.
\end{equation}
It is clear that $ z_1<0$ and $ z_2>0$.  $B$ has to be real-valued.
This is possible for the case $z=z_1$ if $A<0$, and for $z=z_2$ if $A>0$.
 The value of $z_2$ is the same  as that found for the exact soliton solution in \cite[Eq.~(28)]{KRI14}. 
In general 
\begin{equation} \label{bba}
B^2= \frac{A\alpha}{ \beta}z .
\end{equation}
Now, we insert (\ref{bba}) into (\ref{c1}),(\ref{c2}) and (\ref{c3}). Besides a trivial solution with $A=0$ we obtain
\begin{align} \label{c1a}
 1440(1-v) +A\alpha(1-m)\left[1080-135 A\alpha(1-m) \right] \hspace{2ex} &\\
-240A\alpha(1-5m)-30(A\alpha)^2(10-109m+96m^2)\,z &\nonumber \\
 + 4(A\alpha)^2 (19-847m+1159m^2)\, z^2 & = 0, \nonumber 
\\ \label{c2a}
 9 [A\alpha(m-1)+4]+3[A\alpha(75 m-31)-16]\,z\hspace{2ex} & \\
-76 A\alpha(5 m-1)\,z^2 & = 0,\nonumber
\\ \label{c3a}
 9 (3 A\alpha (m-1)+8) +12 (A\alpha(59 m-37)-8)\,z \hspace{2ex} &\\
-608 A\alpha(2 m-1)\,z^2 & = 0.\nonumber 
\end{align}  
From (\ref{c2a}) we find 
\begin{equation} \label{Ac2a}
 A= -\frac{12 (4 z-3)}{\alpha \left[76z^2(5m-1)-z(225m-93)-9(m-1)\right]} 
\end{equation}
but from (\ref{c3a}) it follows that
\begin{equation} \label{Ac3a}
A=  \!-\!\frac{24 (4 z-3)}{\alpha \left[608z^2(2m\!-\!1)\!-\!4z(177m\!-\!111)\!-\!27(m\!-\!1)\right] }. 
\end{equation}
This looks like a contradiction, but substitution
~$z=z_1=(43 - \sqrt{2305})/152$
in both (\ref{Ac2a}) and (\ref{Ac3a}) gives the same result
\begin{equation} \label{Ac23z1}
A_1= \frac{24 \,(71+\sqrt{2305})}{(-329+5\sqrt{2305})\, \alpha \,(m-5)}.
\end{equation}
For   ~$z=z_2=(43 + \sqrt{2305})/152$ 
the common result is
\begin{equation} \label{Ac23z2}
A_2= \frac{24 \,(-71+\sqrt{2305})}{(329+5\sqrt{2305})\, \alpha \,(m-5)}.
\end{equation}
\textcolor{blue}{This means that not only are the equations (\ref{c4}) and (\ref{c5}) equivalent, but also (\ref{c2}) and (\ref{c3}), as well. Therefore the equations (\ref{c2})-(\ref{c5}) supply only three independent conditions for the coefficients of KdV2 solutions in the form (\ref{ey}).
}

Now, using $z=z_1$ and $A_1$ given by (\ref{Ac23z1}) 
 we obtain from (\ref{c1a})
\begin{equation} \label{Vz1}
v_1 = \frac{\text{vnum}_{-}(m) }{\text{vden}_{-}(m)}
\end{equation}
and with  $z=z_2$ and $A_2$ given by (\ref{Ac23z2})
\begin{equation} \label{Vz2} 
v_2 =\frac{\text{vnum}_{+}(m) }{\text{vden}_{+}(m)},
\end{equation}
where
\begin{align} 
\text{vnum}_{\mp}(m) = & ~6 \left\{\left(2912513\mp 58361 \sqrt{2305}\right)
   m^2  \right. \nonumber \\ & \hspace{2ex}
-54 \left(584397\mp 10069 \sqrt{2305}\right)  m \nonumber\\ & \hspace{2ex} \left.
+75245133\mp 1419141 \sqrt{2305}\right\},\nonumber
\end{align}
and 
\begin{equation} 
\text{vden}_{\mp}(m) = 95 \left(329\mp 5 \sqrt{2305}\right)^2 (m-5)^2.\nonumber
\end{equation}

\centerline{{\bf Discussion of mathematical solutions}}
\vspace{2mm}

From a strictly mathematical point of view we found two families of solutions determined by coefficients $A,B,v$ as functions of the elliptic parameter $m$. There are two cases.
\begin{itemize}
\item {\bf Case 1.} $\displaystyle z=z_1=\frac{43-\sqrt{2305}}{152}\approx -0.0329633<0$.  \textcolor{blue}{
This case leads to $B^2<0$ and has to be rejected as in previous papers \cite{KRI14,IKRR17a}. }



\begin{figure}[tbh]
\begin{center}
\resizebox{0.99\columnwidth}{!}{\includegraphics{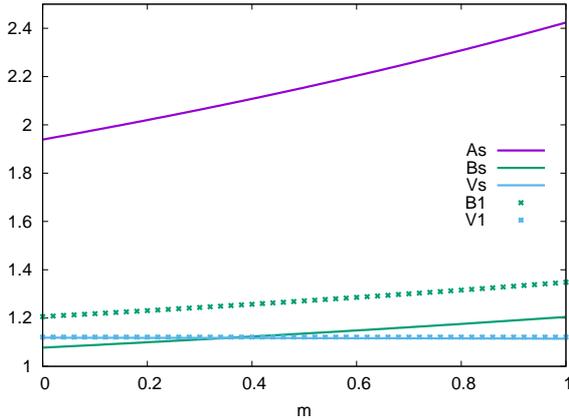}}
\end{center}
\vspace{-5mm}
\caption{The same as in Fig.~\ref{ABVsingle} but for superposition solutions $\eta_+$ (\ref{ey}) and $\eta_-$ (\ref{ey-}).}\label{ABVcndn}
\end{figure}

\item {\bf Case 2.}  $\displaystyle z=z_2=\frac{43+\sqrt{2305}}{152}\approx 0.598753 >0$. Then 
\textcolor{blue}{
\begin{align} \label{A2a}
A &= \frac{12 \left(\sqrt{2305}-51\right)}{37 \alpha  (m-5)} >0,
 \\ \label{B2a}
 B &=  \sqrt{ \frac{12\left(\sqrt{2305}-14\right)}{703(5-m)\beta} }
\end{align}
}
and $v_2$ is given by (\ref{Vz2}). Since $m\in[0,1], ~(m-5)<0$ ~then~
$B_2$~ is real.
The solution in this case is 
\begin{align}  \label{ey2}
\eta_2(x&-v_2t,m)  =  \frac{1}{2} A_2 \left[ \dn^2(B_2 (x-v_2t),m) 
\right. \\ & \left. +\sqrt{m}\,\cn(B_2 (x-v_2t),m)\dn(B_2(x-v_2t),m) \right].\nonumber 
\end{align}
\end{itemize}

Coefficients $A_2, B_2, v_2$ of superposition solutions (\ref{ey}) to KdV2 as functions of $m$ are presented in Fig.~\ref{ABVcndn} for $\alpha=\beta=\frac{1}{10}$ and compared to corresponding  solutions to KdV. Here, similarly as in Fig.~\ref{ABVsingle}, we assume that $A_{\text{KdV}}=A_{\text{KdV2}}$.

\textcolor{blue}{
Physically relevant exact superposition solutions to KdV2 can be found for greater values of the parameter $\alpha$ than $\frac{1}{10}$. In figure 4 
the amplitude $A(\alpha,m)$ given by (\ref{A2a}) is shown as a contour plot for the region $\alpha\in [0.01,0.5], \hspace{1ex} m\in[0,1]$. It is clear that reasonable amplitudes occur in  wide regions of $\alpha$ and $m$ values, similarly like in the case $\dn^2$. The amplitudes become too big only for  $\alpha\to 0$, but in such cases KdV works very well.
\begin{figure}[tbh]
\begin{center}
\resizebox{0.99\columnwidth}{!}{\includegraphics{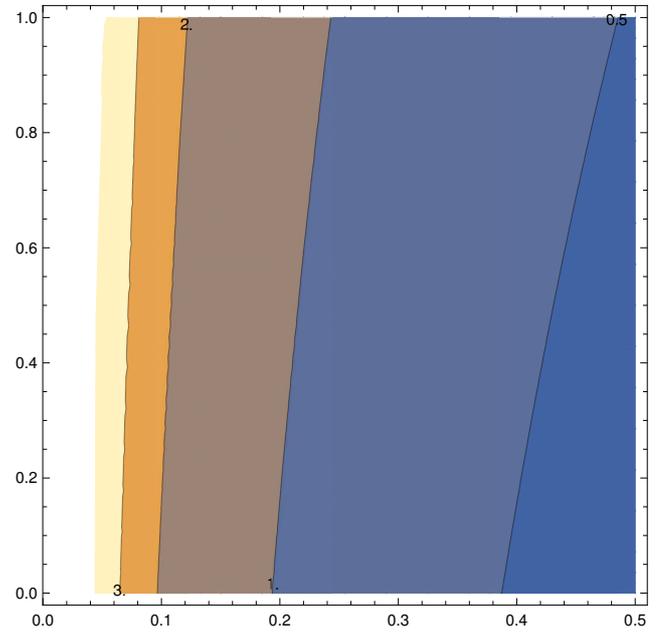}}
\end{center} \textcolor{blue}{
\caption{The same as in figure \ref{Kontur1} but for superposition solutions $\eta_+$.  
} }
\label{Kontur2}\end{figure}
}

\subsection{Superposition "$\dn^2-\sqrt{m}\cn\dn$"} \label{ss2.2}

\begin{figure}[tbh]
\begin{center}
\resizebox{0.99\columnwidth}{!}{\includegraphics{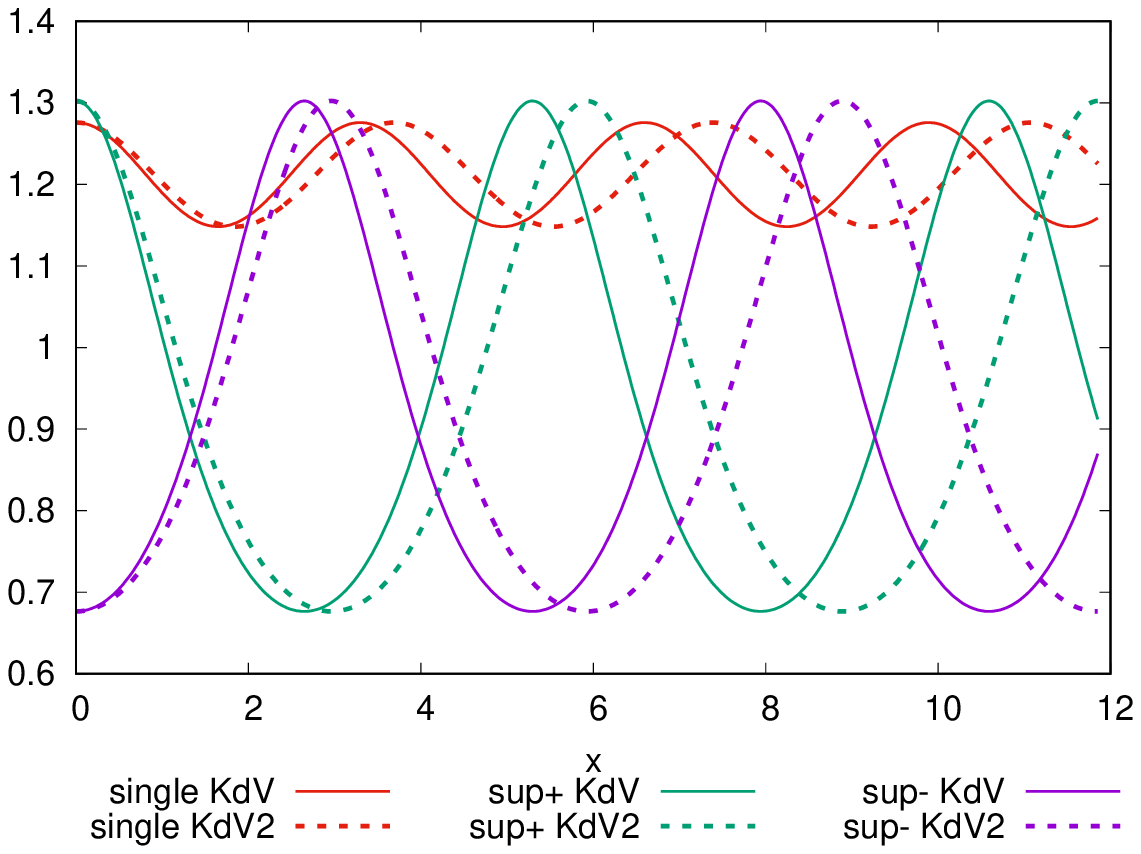}}
\end{center}
\vspace{-5mm}
\caption{Profiles of KdV and KdV2 waves for $m=0.1$.} \label{m01}
\end{figure}

\begin{figure}[tbh]
\begin{center}
\resizebox{0.99\columnwidth}{!}{\includegraphics{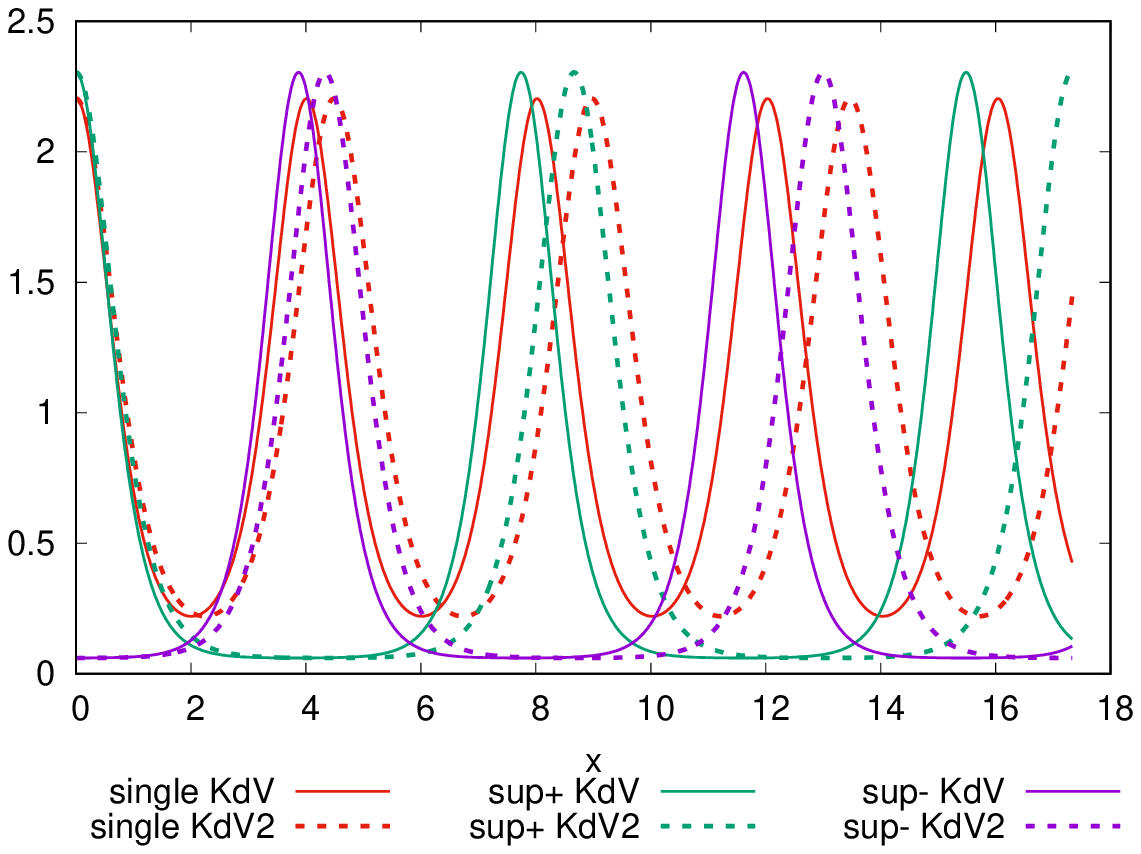}}
\end{center}
\vspace{-5mm}
\caption{Profiles of KdV and KdV2 waves for $m=0.9$.} \label{m09}
\end{figure}

\begin{figure}[tbh]
\begin{center}
\resizebox{0.99\columnwidth}{!}{\includegraphics{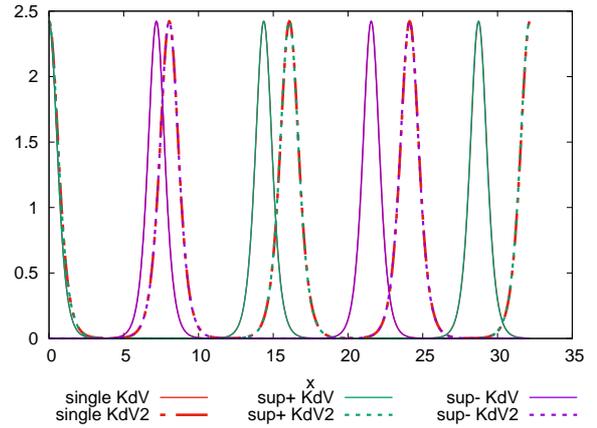}}
\end{center}
\vspace{-5mm}
\caption{Profiles of KdV and KdV2 waves for $m=0.99$.} \label{m99}
\end{figure}

Now we check the alternative superposition "$\dn^2-\sqrt{m}\cn\dn$" 
\begin{equation} \label{ey-}
 \eta_{-}(y) = \frac{1}{2} A \left[ \dn^2(By,m) -\sqrt{m}\,\cn(By,m)\dn(By,m) \right] .
\end{equation}
In this case the derivatives are given by formulas similar to (\ref{y1})-(\ref{y5a}) with some signes altered.
Analogous changes occure in formulas (\ref{E1})-(\ref{E7}).
Then (\ref{eq2}) has a similar form like (\ref{eq2a})
\begin{align}  \label{eq2a-}
 & \frac{1}{2} AB\,\sqrt{m} \left(-\sqrt{m}\cn +\dn\right)^2\sn \, \\ & \hspace{2ex} \times 
\left( F_0 +F_{cd} \cn\dn + F_{c^2} \cn^2 + F_{c^3d}\cn^3\dn+ F_{c^4} \cn^4\right) =0 . \nonumber
\end{align}
Equation (\ref{eq2a-}) is valid for arbitrary arguments  when all  coefficients $F_0,F_{cd},F_{c^2},F_{c^3d}, F_{c^4}$ vanish simultaneously. This gives us a set of  equations for the coefficients $v,A,B$. Despite some changes in signs on the way to (\ref{eq2a-}) this set is  the same as for "$\dn^2+\sqrt{m}\cn\dn$" superposition (\ref{c1})--(\ref{c5}).
Then the coefficients $A,B,v$ for superposition "$\dn^2-\sqrt{m}\cn\dn$" are the same as for  superposition "$\dn^2+\sqrt{m}\cn\dn$" given above. This property for KdV2 is the same as for KdV, see \cite{KhSa}. It follows from periodicity of the Jacobi elliptic functions. From 
$$ 
\cn(y\!+\!2K(m),m)\!=-\cn(y,m),\hspace{1ex}   \dn(y\!+\!2K(m),m)\!=\!\dn(y,m)$$
it follows that
\begin{align}  \label{perJ}
\dn^2(y&\!+\!2K(m),m)\!+\!\sqrt(m)\cn(x\!+\!2K(m),m)\dn(x\!+\!2K(m),m) \nonumber\\ &
= \dn^2(x,m)-\sqrt(m)\cn(x,m)\dn(x,m). 
\end{align}
So both superpositions $\eta_{+}$ (\ref{ey}) and $\eta_{-}$ (\ref{ey-}) represent the same solution, but shifted by the period of the Jacobi elliptic functions.
This property is well seen in figures \ref{m01}-\ref{m99}.

\section{Examples}

Below, some examples of wave profiles for both KdV and KdV2 are presented. 
We know from section \ref{sec1} that for a given $m$, the coefficients $A,B,v$ of KdV2 solutions are fixed. As we have already written, this is not the case for $A,B,v$ of KdV solutions. So, there is one free parameter.
In order to compare KdV2 solutions to those of KdV for identical $m$, we set $A_{\text{KdV}} = A_{\text{KdV2}}$. In figures \ref{m01}-\ref{m99} below, KdV solutions of the forms (\ref{adn2}), (\ref{ey}) and (\ref{ey-})  are drawn with solid  red, green and blue lines, respectively. 
For KdV2 solutions the same color convention  is used, but with dashed lines. 
\textcolor{blue}{In all the presented cases the parameters $\alpha=\beta=0.1$ were used.}

Comparison of wave profiles for different $m$ suggests several observations.
For small $m$, solutions given by the single formula (\ref{adn2}) differ substantially from those given by superpositions (\ref{ey}) and (\ref{ey-}). Note that (\ref{adn2}) is equal to the sum of both superpositions and when $m\to 1$ the distance between crests of $\eta_{+}$ and $\eta_{-}$ increases to infinity (in the $m=1$ limit). All three solutions converge to the same soliton.

\textcolor{blue}{
In order to check whether the obtained analytic solutions are really true solutions to KdV2 several numerical simulations were performed. In each of them the numerical FDM  code used with success in previous studies \cite{KRR14,KRI14,KRI15,KRIR17,IKRR17a,IKRR17b} was applied. Since the calculations concerned periodic solutions the periodic boundary conditions were used with an $x$ interval equal to the particular wavelength. In figures \ref{m01}-\ref{m99} dashed lines display profiles of single $\dn^2$ (\ref{adn2}) and superposition $\eta_+$ and $\eta_-$ (\ref{adncn}) solutions for three values of $m=0.1, 0.9$ and 0.99.
Below in figures 6-8 
six examples of time evolution for these solutions obtained in numerics are presented. Profiles of solutions at  time instants $t=0, T/4, T/2, 3T/4$ and $T$, where $ T=\lambda/v$ are displayed.
Open symbols represent the profiles at $t=T$ which overlap with those at $t=0$
with numerical deviations less than 10$^{-11}$. In all the presented examples, as well as all others not shown here, numerics confirmed a uniform motion and a fixed shape for the considered solutions. 
}

\begin{figure}[tbh]
\begin{center}
\resizebox{0.99\columnwidth}{!}{\includegraphics[angle=270]{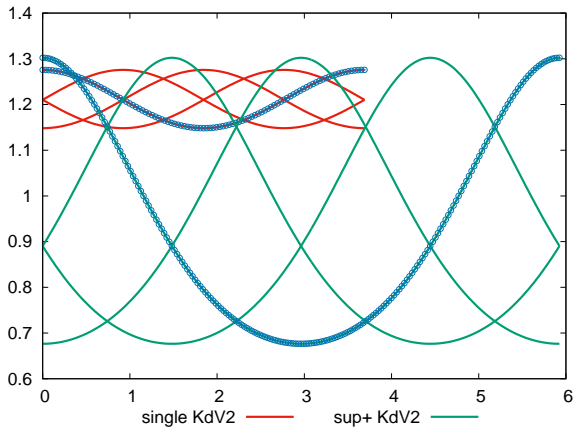}}
\end{center}  \textcolor{blue}{
\caption{Time evolution of single KdV2 solution (red lines) and $\eta_+$   KdV2 solution for $m=0.1$. }  }
\label{m01num}
\end{figure}

\begin{figure}[tbh]
\begin{center}
\resizebox{0.99\columnwidth}{!}{\includegraphics[angle=270]{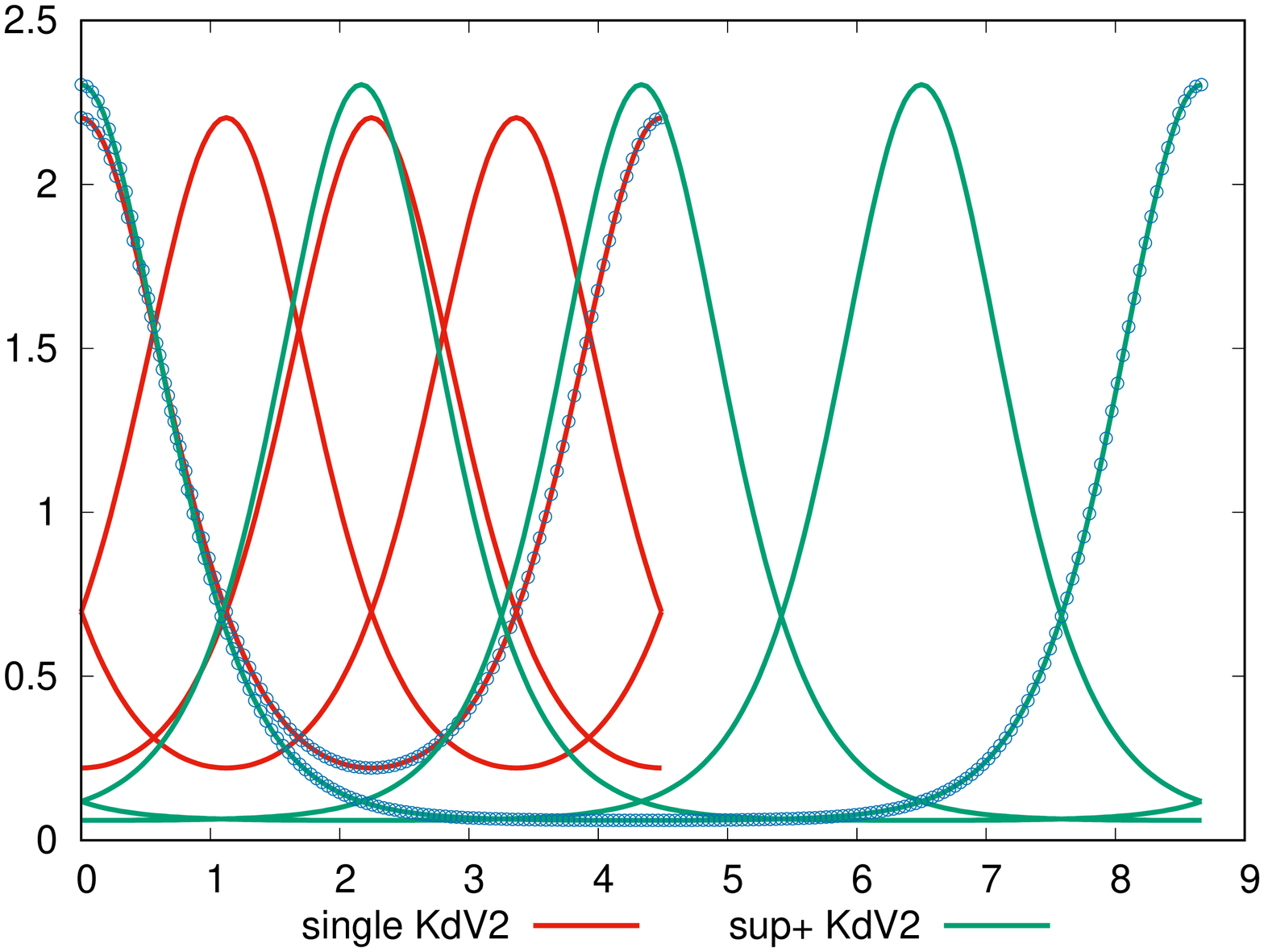}}
\end{center}  \textcolor{blue}{
\caption{The same as in figure 6 
but for $m=0.9$. }  }
\label{m09num}
\end{figure}

\begin{figure}[tbh] 
\begin{center}
\resizebox{0.99\columnwidth}{!}{\includegraphics[angle=270]{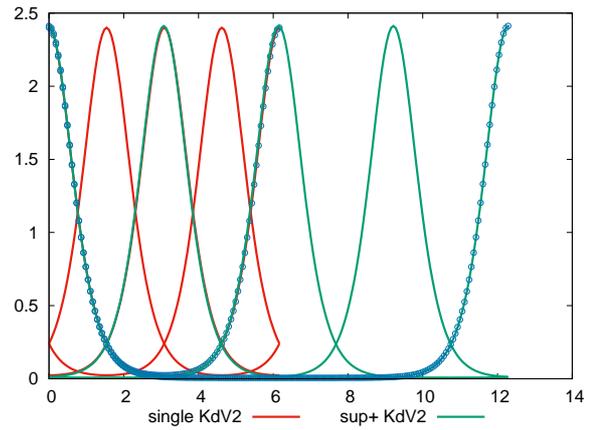}}
\end{center}  \textcolor{blue}{
\caption{The same as in figure 6 
but for $m=0.99$. }  }
\label{m99num}
\end{figure}

\section{Conclusions} 

The most important results of the paper can be summarized  as follows.

 It is shown that
several kinds of analytic solutions of KdV2 have the same forms as 
corresponding solutions to KdV but with different coefficients.
This statement is true for our single solitonic solutions \cite{KRI14}, periodic solutions in the form of single Jacobi elliptic functions $\cn^2$ \cite{IKRR17a} or $\dn^2$, and for periodic solutions in the form of superpositions $\dn^2\pm\sqrt{m}\cn\dn$ (this paper).
Coefficients $A,B,v$ of these solutions to KdV2 are fixed by coefficients of the equation, that is by values of $\alpha,\beta$ parameters. This is in contradiction to the KdV case where one coefficient (usually $A$) is arbitrary.  



\end{document}